\documentclass[12pt]{article}
\usepackage{amssymb,amsmath,epsfig}
\begin{document}
 \title{\bf Interior Solutions  of Fluid Sphere in $f(R,T)$ Gravity Admitting Conformal Killing Vectors}

 \author{M. Zubair \thanks{mzubairkk@gmail.com; drmzubair@ciitlahore.edu.pk} ${}^{(a)}$,
I. H. Sardar \thanks{iftikar.spm@gmail.com} ${}^{(b)}$, F.
Rahaman${}^{(b)}$ \thanks{rahaman@associates.iucaa.in}and G.
Abbas$^{(c)}$ \thanks{abbasg91@yahoo.com} \\\
 ${}^{(a)}$ Department of Mathematics,\\ COMSATS Institute of
 Information Technology, Lahore, Pakistan.\\ ${}^{(b)}$ Department of
 Mathematics,\\Jadavpur University , Kolkata - 700032
, India.\\ ${}^{(c)}$ Department of Mathematics,\\ The Islamia
University of Bahawalpur, Bahawalpur, Pakistan.}
\date{}
\maketitle
 \begin{abstract}

We discuss the interior solutions of fluid Sphere in $f(R,T)$
gravity admitting conformal killing vectors, where $R$ is Ricci
scalar and $T$ is trace of energy momentum tensor. The solutions
corresponding to isotropic and anisotropic configurations have been
investigated explicitly. Further, the anisotropic case has been
dealt by the utilization of linear equation of state. The results
for both cases have been interpreted graphically. The equation of
state parameter, integration constants and other parameters of the
theory have been chosen to find the central density equal to
standard value of central density of the compact objects. The energy
conditions as well as stability of the solutions have been
investigated in the background of $f(R,T)$ gravity.
\end{abstract}


 {\bf Keywords:} Compact Stars, $f(R, T)$ Gravity.\\

 {\bf PACS:} 04.20.Cv; 04.20.Dw

\section{Introduction}

It is an admitted fact that the accelerated expansion of our
Universe and the existence of dark matter are two such important
aspects of modern cosmology that have been accepted on the
background of observational data (Riess (2007),   Perlmutter (1999),  Hanany (2000),   Peebles  and Ratra (2003) ). These finding
have imposed some additional challenges to theories of gravitation.
The most significant way to explain the observational data is by
admitting that the Einstein theory of gravitation breaks down at
large scales, and a more generalized form of action is required to
describe the gravitational field at large scales. During the last
decades the most general theoretical models of $f(R)$ gravity, where
$R$ being Ricci scalar , have been extensively used to explain the
cosmological results. The accelerated expansion of universe and the
conditions for the presence of dark energy have been have studied in
$f(R)$ gravity ( Padmanabhan (2003)). The physical conditions for the viable
cosmological models have been found in $f(R)$ (   Nojiri  and Odintsov (2011),
 Bamba et al(2012), and
satisfy the weak field limit obtained from the classical tests of
general relativity. The $f(R)$ models that satisfy the solar system
tests of general relativity and provides the unification of
inflation and dark energy were investigated in (  Nojiri  and Odintsov  (2007) ,    Cognola et al ,  (2008)).
In $f(R)$ gravity, it has been proved that the galactic dynamic of
massive test particles can be explained by excluding the possibility
of dark matter ([  Capozziello et al  (2006) ,    Borowiec et al  (2007) ,  Martins   and Salucci (2007) ,  Boehmer et al ,   (2008)). Further, investigations in
$f(R)$ gravity can be found in detail in ([   Sotiriou  and Faraoni (2010)   Lobo (2008) ,    Capozziello and V. Faraoni  (2010)).

A most general form of $f(R)$ theory of gravity was proposed in
(   Bertolami et al  (2007) ), by including an arbitrary function of the Ricci scalar
$R$ with the matter Lagrangian density $L_m$ in the action of the
theory. As a consequence of such modification the motion of massive
particles is non-geodesic and there exists an extra-force
 The astronomical implication of non-minimal matter-geometry
 coupling were explored in (  Nojiri and Odintsov (2004) ,  Harko (2010) ) and Palatini
approach of non-minimal geometry-coupling models was discussed in
(  Harko   and Lobo (2010) ). In this coupling, a maximal extension of the
Hilbert-Einstein action was performed in (Koivisto (2006))  by taking the
gravitational Lagrangian as an arbitrary function of Ricci scalar
$R$ and matter Lagrangian density $L_m$.

The field equations as well as the equations of motion for test
particles have been formulated in the metric formalism, which is the
covariant divergence of the stress-energy tensor. A specific form of
above coupling was considered as another extension of general
relativity as $f(R,T)$,  modified theories of gravity, where action
is given by an arbitrary function of the Ricci scalar $R$ and trace
of the stress-energy tensor $T$ (Nesseris (2009)). Firstly, Lobo, et. al.
(  Harko et al(2011)) introduced such modifications to obtain some specific results
of cosmology, the more general aspects such as reconstruction of
cosmological models and late time acceleration of universe was first
studied in ( Houndjo (2012)). Further, the energy conditions and
thermodynamics in $f(R,T)$, theories have been investigated by
Sharif and Zubair ( Sharif  and Zubair (2012) ,  Sharif, M. and Zubair (2012)).

In general conformal Killing vectors (CKVs)explain the mathematical
relation between the geometry and contents of matter in the
spacetime via Einstein set of field equations. The CKVs are used
generate the exact solution of the Einstein field equation in more
convenient form as compared to other analytical approaches. Further
these are used to discover the conservation laws in any spacetime.
The Einstein field equations being the highly non-linear partial
differential equations can be reduced to a set of ordinary
differential equations by using CKVs. A lot of astrophysical
phenomena have been explored on the theoretical background using the
CKVs approach (see (  Ray et al  (2008),  Rahaman et al (2014),   Rahaman et al  (2015a,b,c)  ). The interior anisotropic
fluid spheres admitting conformal motion have been studied during
the last stages by Herrera and his collaborators
(  Herrera (1992)  Herrera et al (1984),  Herrera and  Ponce de Leon (1985)   Herrera and  Ponce de Leon (1985a,b) 
 ).

In the present paper, our main motivation is to find the exact
solution for static anisotropic spheres preserving the conformal
motion in $f(R,T)$ gravity. Section \textbf{2} deals with
formulation of field equations in $f(R,T)$ gravity. The exact
solutions with isotropic and anisotropic configurations have been
investigated in section \textbf{3}. The last section summaries the
results of the paper.

\section{ Interior Matter Distribution in $f(R, T)$ Gravity}

The modified action in $f(R,T)$ is as follows  
\begin{equation}\label{1}
\int dx^4\sqrt{-g}[\frac{f(R, T)}{16\pi G}+\mathcal{L} _ {(m)}],
\end{equation}
where $\mathcal{L} _ {(m)}$ is matter Lagrangian and $g$ denote the
metric tensor. Different choices of $\mathcal{L} _ {(m)}$ can be
considered, each of which directs to a specific form of fluid. The
line element for general spherically symmetric metric describing the
compact star stellar configuration is
\begin{equation}\label{2}
ds^2=e^{a(r)}dt^{2}-e^{b(r)}dr^{2}-r^2(d\theta^{2}+\sin^{2}\theta
d\phi^{2}).
\end{equation}

Taking $8\pi G = 1$ and upon variation of modified EH action in
$f(R,T)$ (\ref{1}) with respect to metric tensor $g_{uv}$, the
following modified field equations are formed as
\begin{eqnarray}\nonumber
G_{uv}&=&\frac{1}{f_R}\left[(f_T+1)T^{(m)}_{uv}-\rho g_{uv}f_T+
\frac{f-Rf_R}{2}g_{uv}\right.\\\label{3}&+&\left.(\nabla_u\nabla_v-g_{uv}\Box)f_R\right],
\end{eqnarray}
where $T^{(m)}_{uv}$ denotes the usual matter energy momentum tensor
that is considered to be anisotropic, is given by
\begin{equation}\label{4}
T^{(m)}_{uv}=(\rho+p_{t})V_{u}V_{v}-p_{t}g_{uv}+(p_{r}-p_{t})\chi_{u}\chi_{v},
\end{equation}
where $\rho$, $p_r$ and $p_t$ denote energy density, radial and
transverse stresses respectively. The four velocity is denoted by
$V_{u}$ and $\chi_{u}$ to be the radial four vector satisfying
\begin{equation}\label{5}
V^{u}=e^{\frac{-a}{2}}\delta^{u}_{0},\quad V^{u}V_{u}=1,\quad
\chi^{u}=e^{\frac{-b}{2}}\delta^u_1,\quad \chi^{u}\chi_{u}=-1.
\end{equation}
The conformal Killing vector is defined through the relation
\begin{equation}\label{6}
\mathcal{L}_\xi{g}_{\mu\nu}=g_{\eta\nu}\xi^{\eta}_{;\mu}+g_{\mu\eta}\xi^\eta_{;\nu}=\psi(r)g_{\mu\nu},
\end{equation}
where $\mathcal{L}$ represents the Lie derivative of metric tensor
and $\psi(r)$ is the conformal vector.

Using Eq.(\ref{2}) in (\ref{6}), one can find \cite{29a}
\begin{eqnarray}\nonumber
\xi^1a'=\psi,\\\nonumber \xi^1=\frac{\xi{r}}{2},\\\nonumber
\xi^1b'+2\xi^1_{,1}=\psi,
\end{eqnarray}
These results imply
\begin{eqnarray}\nonumber
e^{a}=C_1r^2, \\\label{7} e^{b}=\left(\frac{C_2}{\psi}\right)^2,
\end{eqnarray}
where $C_1$ and $C_2$ are integration constants.

When $f(R, T)=f_1(R) +\lambda T$, the expression for $\rho$, $p_r$
and $p_t$ can be extracted from modified field equations as follows
\begin{eqnarray}\nonumber
\rho&=&\frac{e^{-b(r)}}{4r^2(1+\lambda)(1+2\lambda)}\{-2(2f_{1R}(-2+(-5+e^{b(r)}\lambda)+r(f'_{1R}(4+3\lambda)
\\\label{8}&+&r(e^{b(r)}f(1+\lambda)+f''_{1R}(2+3\lambda))))+r(-2f_{1R}+f'_{1R}r(2+3\lambda))b'(r)\},\\\nonumber
p_r&=&\frac{e^{-b(r)}}{4r^2(1+\lambda)(1+2\lambda)}\{2(-(2f_{1R}+f''_{1R}r^2)\lambda+f'_{1R}r(6+7\lambda)
+e^{b(r)}(2f_{1R}\lambda\\\label{9}&+&fr^2(1+\lambda))+r(f'_{1R}r\lambda+f_{1R}(6+8\lambda))b'(r)\},\\\nonumber
p_t&=&\frac{e^{-b(r)}}{4r^2(1+\lambda)(1+2\lambda)}\{2(2f_{1R}(-2-3\lambda+e^{b(r)}(1+\lambda))+r(f'_{1R}(4+9\lambda)
\\\nonumber&+&r(e^{b(r)}f(1+2\lambda)+f''_{1R}(2+3\lambda))))+r(-f'_{1R}r(2+3\lambda)+f_{1R}(2\\\label{10}&+&6\lambda))b'(r)\}.
\end{eqnarray}
Here $f_{1R}=\frac{df_1}{dR}$ and prime denotes the derivatives with
respect to radial coordinate. Eqs.(\ref{8})-(\ref{10}) are highly
non-linear to find the $e^{b(r)}$. Therefore, we consider the simples case $f(R,T)=R+\lambda T$
which represents the $\Lambda$CDM model in $f(R,T)$ gravity. For this choice we can find the results for
$\rho$, $p_r$ and $p_t$ in the following form
\begin{eqnarray}\label{11}
\rho&=&\frac{e^{-b(r)}}{2r^2(1+\lambda)(1+2\lambda)}\{2(-1+e^{b(R)}+2\lambda)+r(2+3\lambda)b'(r)\},\\\label{12}
p_r&=&\frac{e^{-b(r)}}{4r^2(1+\lambda)(1+2\lambda)}\{6-2e^{b(r)}+4\lambda+r\lambda b'(r)\},\\\label{13}
p_t&=&\frac{e^{-b(r)}}{4r^2(1+\lambda)(1+2\lambda)}\{2-2(-3+e^{b(r)})\lambda-r(2+3\lambda)b'(r)\}.
\end{eqnarray}

\section { Solutions }

Now we are seeking solutions for two different physical situations. At first, we assume the isotropic case and secondly we will consider anisotropic model of the Fluid Sphere.

\subsection{Isotropic case }

For isotropic model of the Fluid Sphere, it is assumed that
 $p_r=p_t=p$.

Using the  isotropic pressures and solving  the equations (\ref{11})-(\ref{13}), we get
\begin{eqnarray}\label{14}
e^{-b(r)}&=&\frac{1-\lambda}{2-\lambda}+C_3r^{2-\lambda/2\lambda+1},\\\label{15}
R&=&\frac{1}{r^2}\{4+\frac{6}{\lambda-2}+\frac{3C_3(4+3\lambda)r^{2-\lambda/2\lambda+1}}
{1+2\lambda}\},\\\nonumber
\rho&=&\frac{r^{-2-\frac{\lambda}{1+2\lambda}}}{2(\lambda-2)(\lambda+1)(2\lambda+1)^2}\{2r^{\lambda/1+2\lambda}
(-1-4\lambda-2\lambda^2+4\lambda^3)\\\label{16}&+&C_3r^{2/1+2\lambda}(12+2\lambda-26\lambda^2+11\lambda^3)\},\\\nonumber
p&=&\frac{r^{-2-\frac{\lambda}{1+2\lambda}}}{2(\lambda-2)(\lambda+1)(2\lambda+1)^2}\{2r^{\lambda/1+2\lambda}
(-1-2\lambda+2\lambda^2+4\lambda^3)\\\label{17}&+&C_3r^{2/1+2\lambda}(-12-22\lambda-4\lambda^2+9\lambda^3)\},\\\label{18}
\psi&=&C_2(\frac{1-\lambda}{2-\lambda}+C_3r^{2-\lambda/2\lambda+1})^{1/2},
\end{eqnarray}
where $C_3$ is an arbitrary constant.

To search the physical properties of the interior of the fluid
sphere, we draw the profile of matter density   and   pressure    in
fig.\textbf{1}(left) and fig.\textbf{1} (midle) respectively.  The
profile indicates that  matter density   and   pressure  all are
positive inside the fluid Sphere.  It is to be  noted that
  density and radial pressure are decreasing with the radial coordinate $r$ which are the common features.
  Obviously all energy conditions are satisfied see fig.\textbf{1} (right).   Here, the model indicates  equation of
  state parameter as well as sound velocity  are less than unity, see fig \textbf{2}. Thus our solutions satisfy all
  criteria for physically valid solution of a fluid sphere.

\begin{figure}[thbp]
\centering
\includegraphics[width=0.32\textwidth]{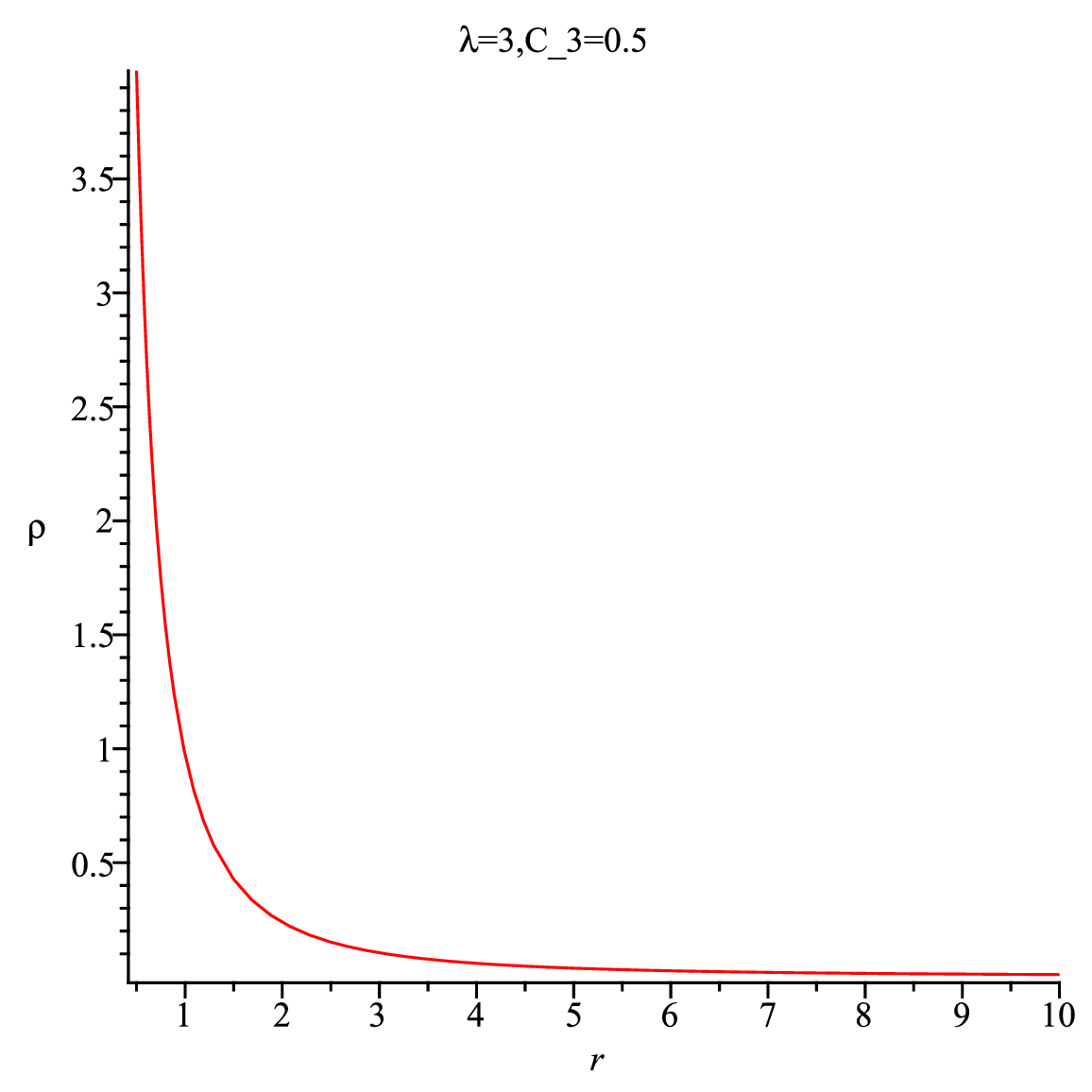}
\includegraphics[width=0.32\textwidth]{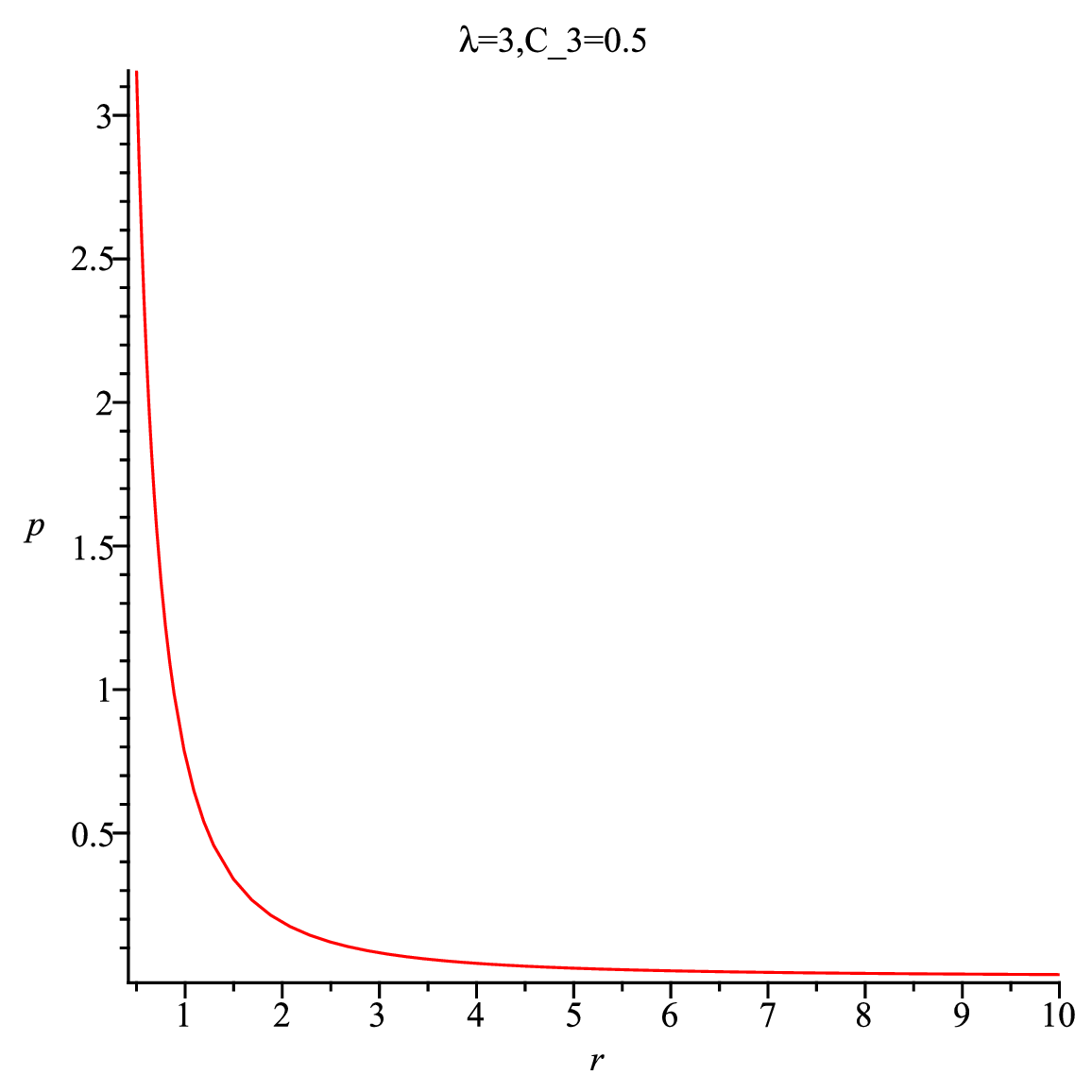}
\includegraphics[width=0.32\textwidth]{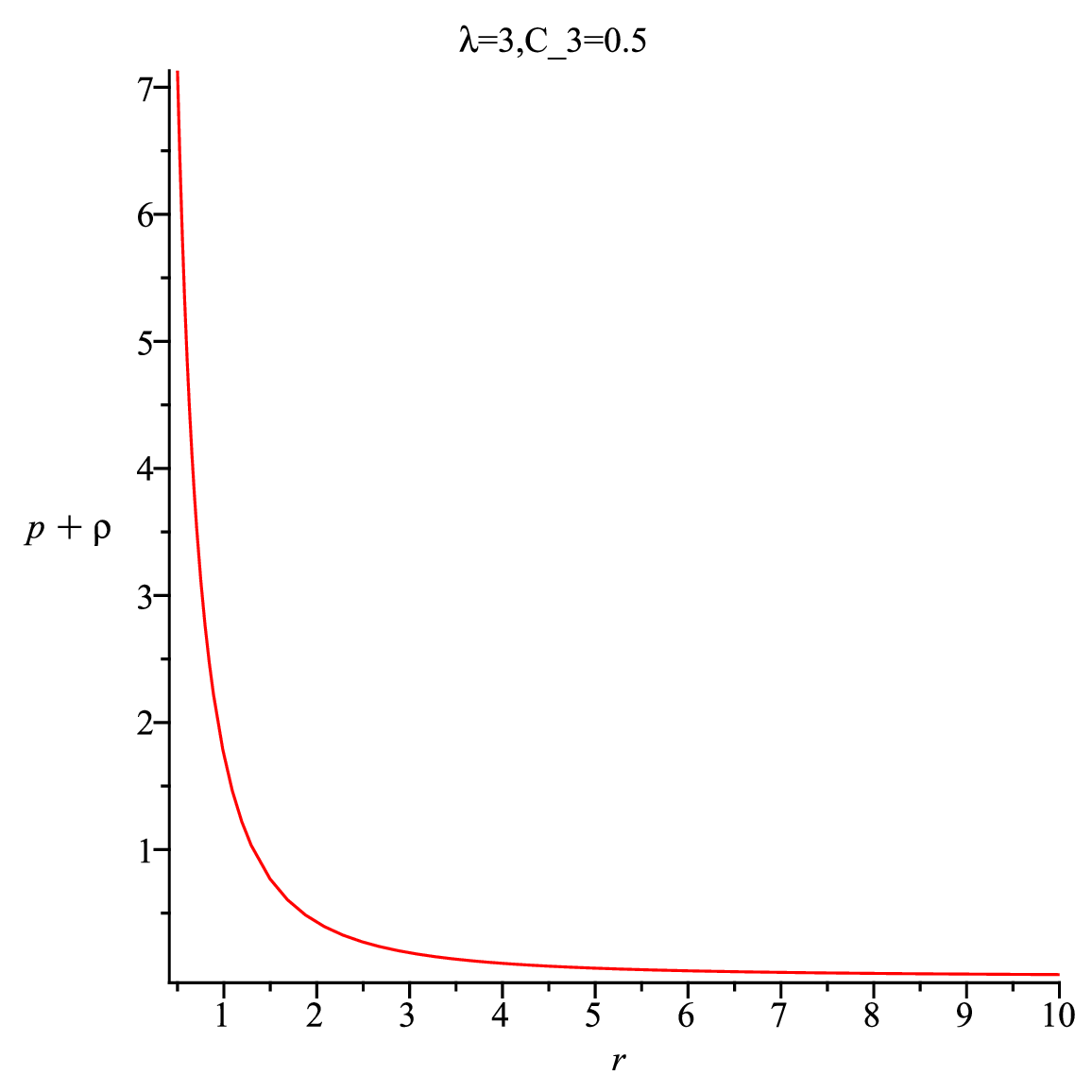}
\caption{ (  left) Density    is  plotted against $r$.
  ( middle)   Pressure  is  plotted against $r$.
  (  right)  Variation of $p+\rho$ is  shown  against $r$.}
\end{figure}

\begin{figure}[thbp]
\centering
\includegraphics[width=0.4\textwidth]{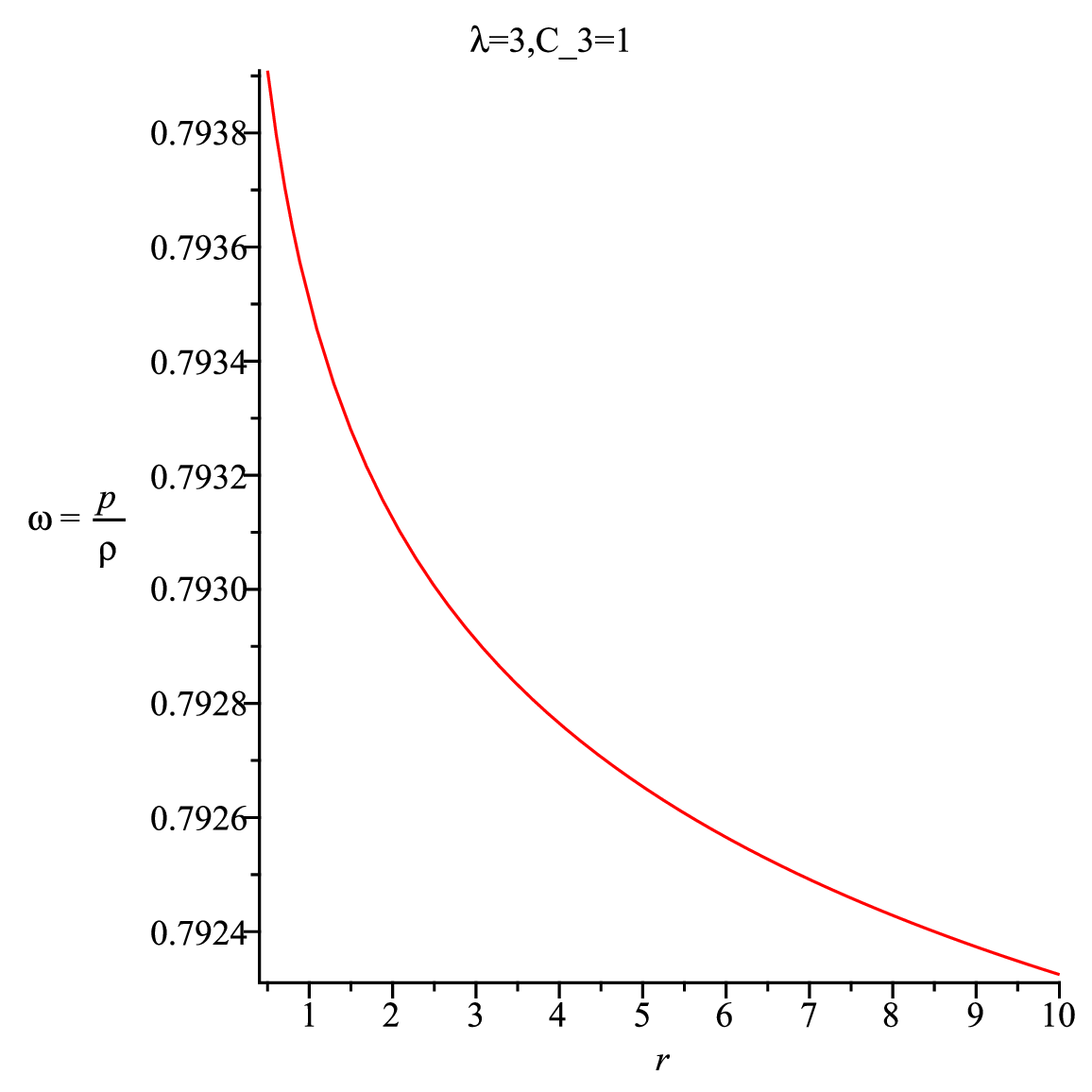}
\includegraphics[width=0.4\textwidth]{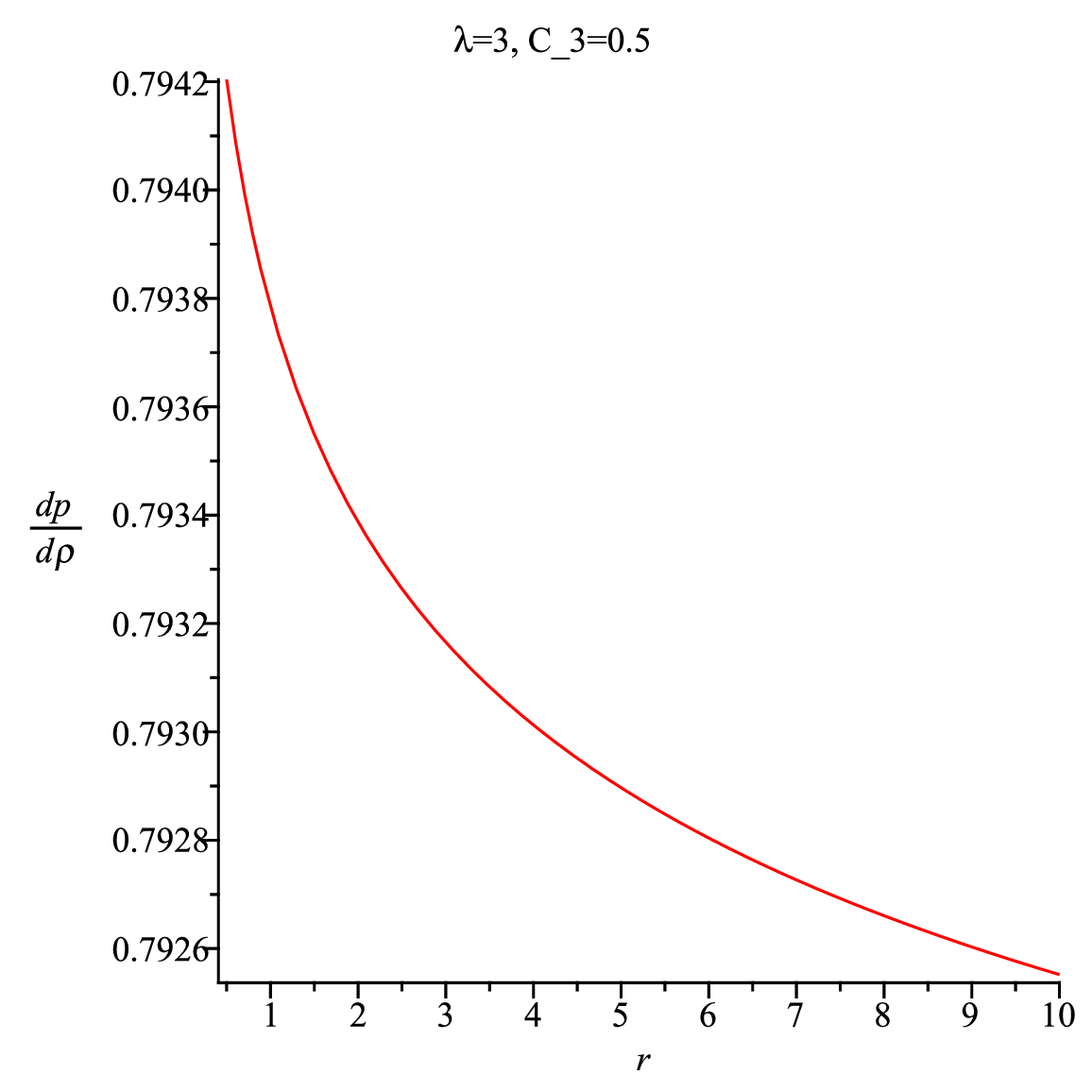}
\caption{ (  left) EoS     is  plotted against $r$.
    (  right)  Variation of sound speed
     is  shown  against $r$.}
\end{figure}

\subsection{Anisotropic case }

Our objective in this section is to develop a model for anisotropic fluid and, therefore, we
assume $p_r\neq p_t$.  The simplest form of the fluid sphere  EoS having the form
\begin{equation}\label{19}
p_r=\omega \rho.
\end{equation}
Therefore the solutions are obtained in the following form
\begin{eqnarray}\label{20}
e^{-b(r)}&=&\frac{2(1+\omega)}{(6+4\lambda)+\omega(2-4\lambda)}
+C_4r^\frac{2(1+\omega)}{\lambda-\omega(2+3\lambda)},\\\label{21}
R&=&\frac{4((1+\lambda)\omega-\lambda)}{r^2(3+\omega+(1-\omega)2\lambda)}
+\frac{C_46((1+3\lambda)\omega-(1+\lambda))r^{-2-\frac{2(1+\omega)}{-\lambda+\omega(2+3\lambda)}}}
{-\lambda+\omega(2+3\lambda)},\\\nonumber
\rho&=&\frac{1}{r^2(1+\lambda)(1+2\lambda)}\{
\frac{2C_4 (1+\lambda(2-\lambda+\omega(2+3\lambda)))r^{-\frac{2(1+\omega)}{-\lambda+\omega(2+3\lambda)}}}
{-\lambda+\omega(2+3\lambda)}\\\label{22}&+&\frac{2+4\lambda}{3+\omega+2\lambda(1-\omega)}\},\\\nonumber
p_r&=&\frac{1}{r^2(1+\lambda)(1+2\lambda)}\{
\frac{2C_4(-\lambda(1+\lambda)+\omega(3+\lambda(7+3\lambda)))r^{-\frac{2(1+\omega)}{-\lambda+\omega(2+3\lambda)}}}
{-\lambda+\omega(2+3\lambda)}\\\label{23}&+&\frac{2\omega(1+2\lambda)}{3+\omega+2\lambda(1-\omega)}\},\\\nonumber
p_r&=&\frac{1}{r^2(1+\lambda)(1+2\lambda)}\{
\frac{C_4(-2+\lambda(-4-3\lambda+\omega(6+9\lambda)))r^{-\frac{2(1+\omega)}{-\lambda+\omega(2+3\lambda)}}}
{-\lambda+\omega(2+3\lambda)}\\\label{24}&+&\frac{1+\omega+2\omega\lambda+2(-1+\omega)\lambda^2}
{3+\omega+2\lambda(1-\omega)}\},\\\label{25}
\psi&=&C_2\left(\frac{2(1+\omega)}{(6+4\lambda)+\omega(2-4\lambda)}
+C_4r^\frac{2(1+\omega)}{\lambda-\omega(2+3\lambda)}\right)^{1/2}.
\end{eqnarray}

\begin{figure}[thbp]
\centering
\includegraphics[width=0.325\textwidth]{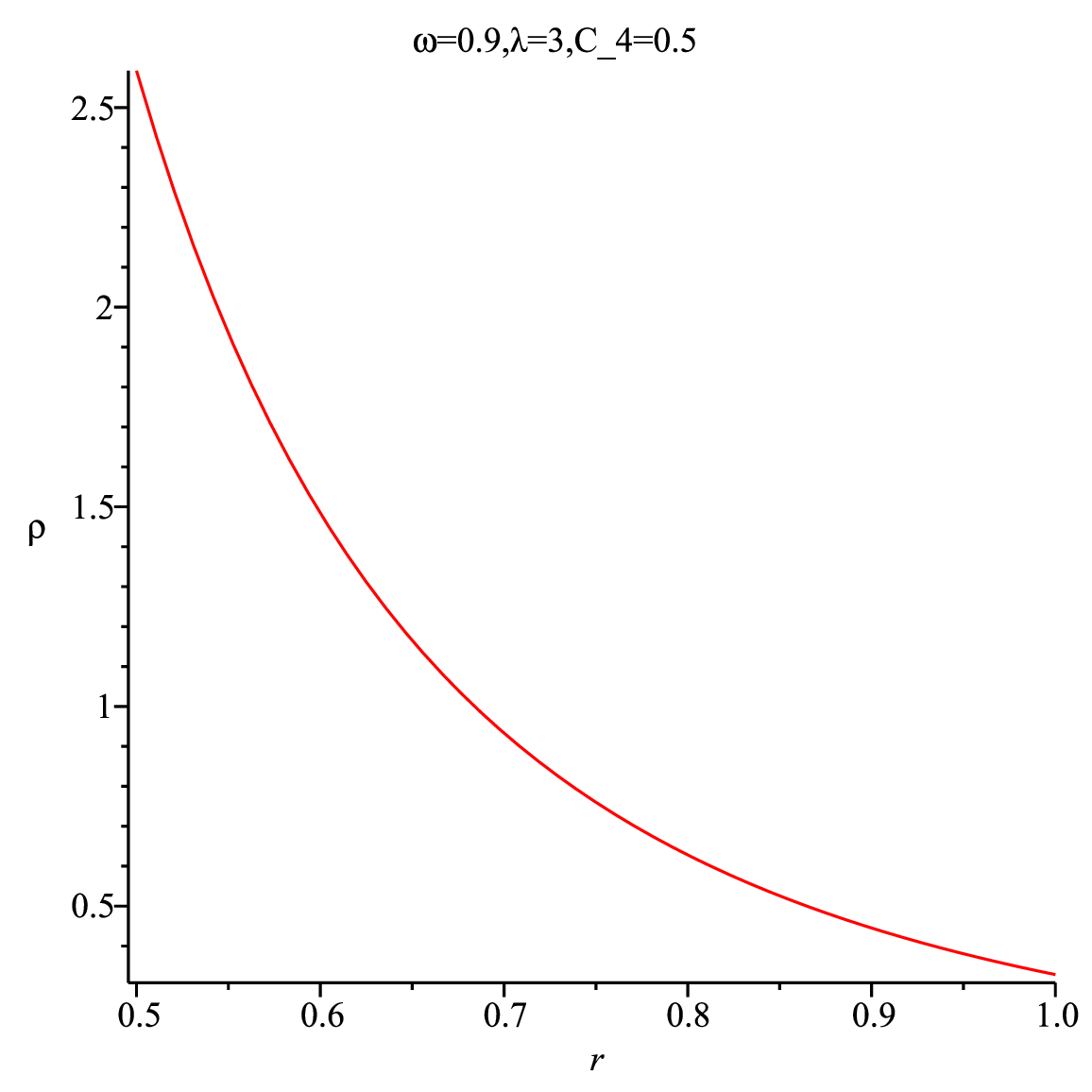}
\includegraphics[width=0.32\textwidth]{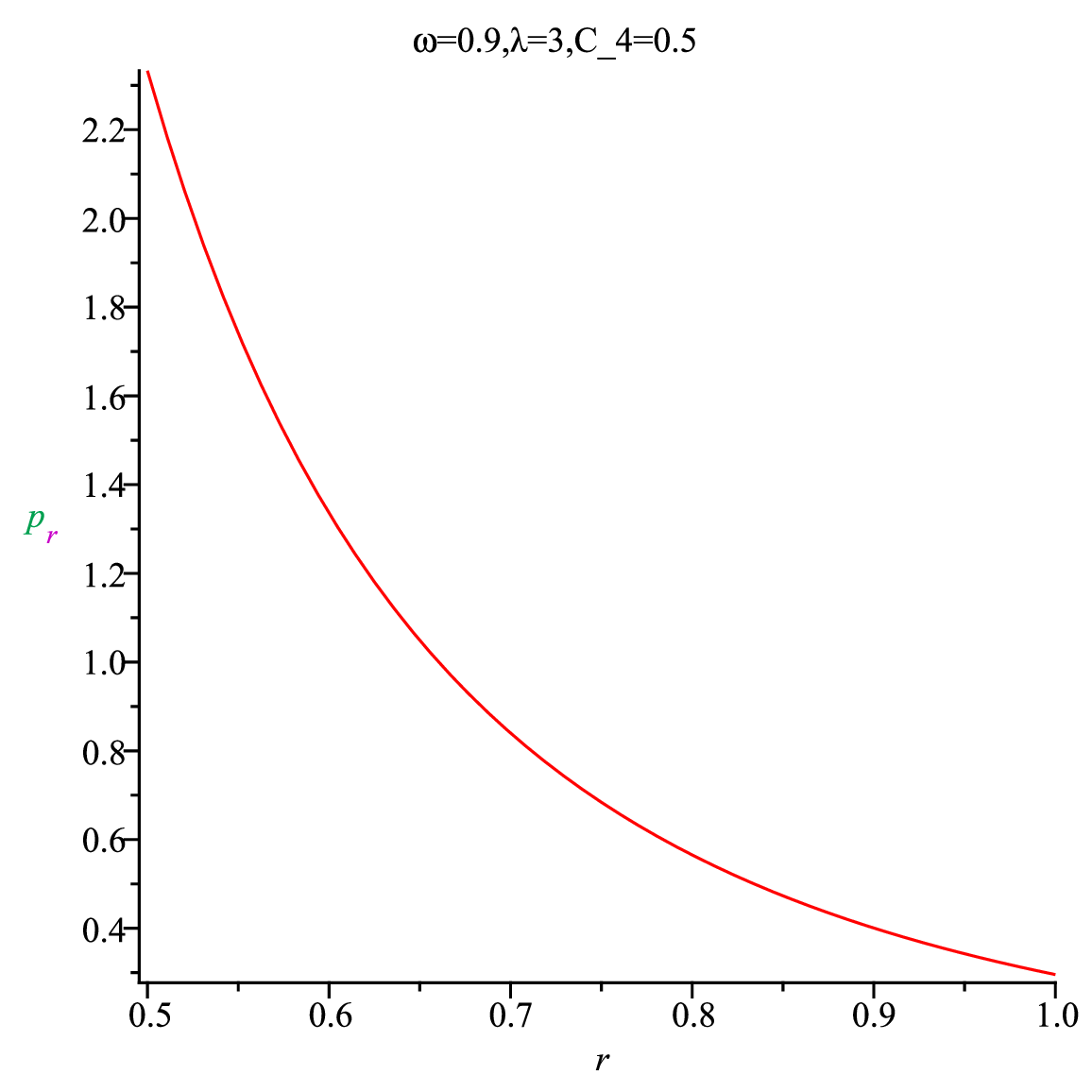}
\includegraphics[width=0.32\textwidth]{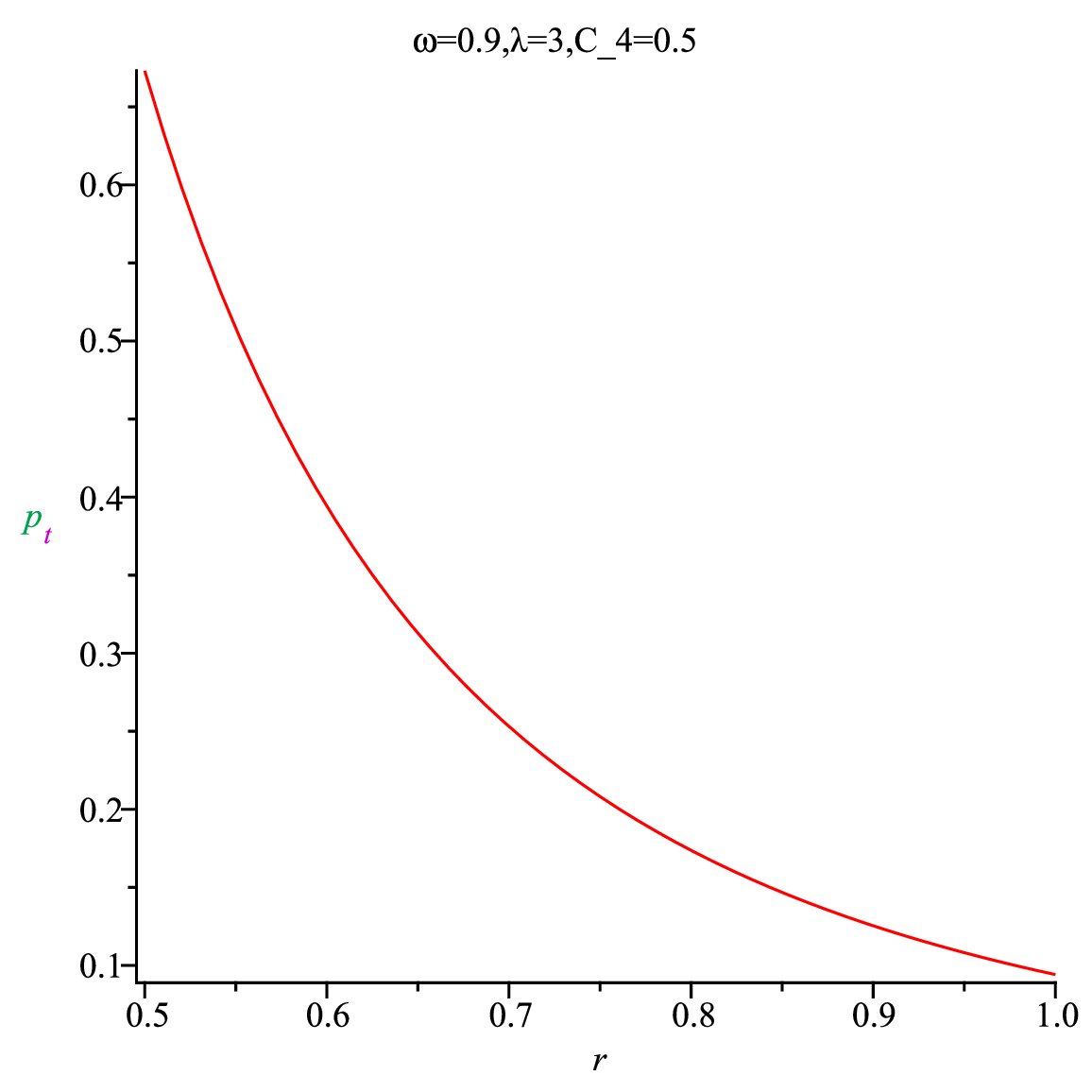}
\caption{ (  left) Density    is  plotted against $r$.
  ( middle)   Radial pressure  is  plotted against $r$.
  (  right)  Transverse pressure  is  plotted against $r$.}
\end{figure}

 \begin{figure}[thbp]
\centering
\includegraphics[width=0.32\textwidth]{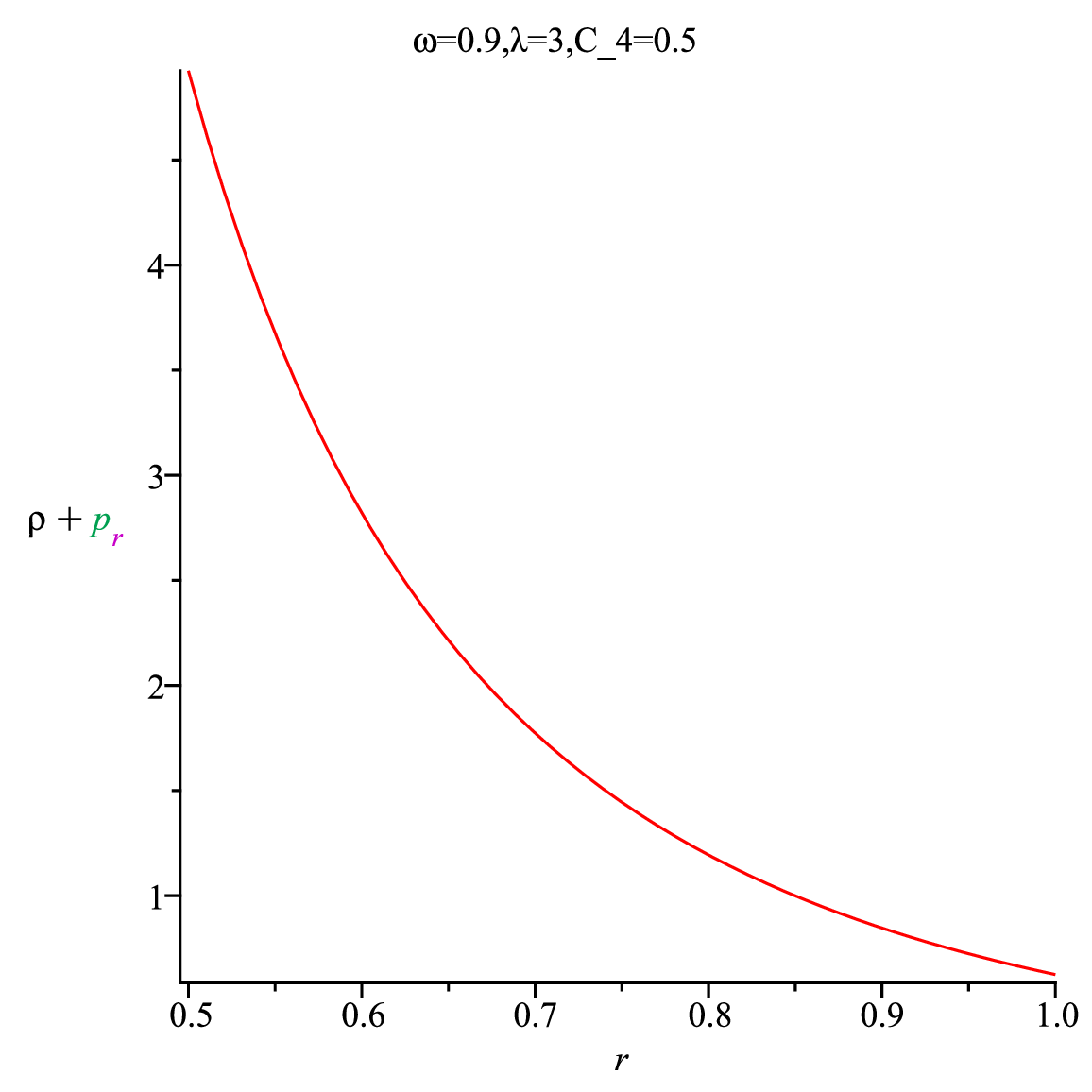}
\includegraphics[width=0.32\textwidth]{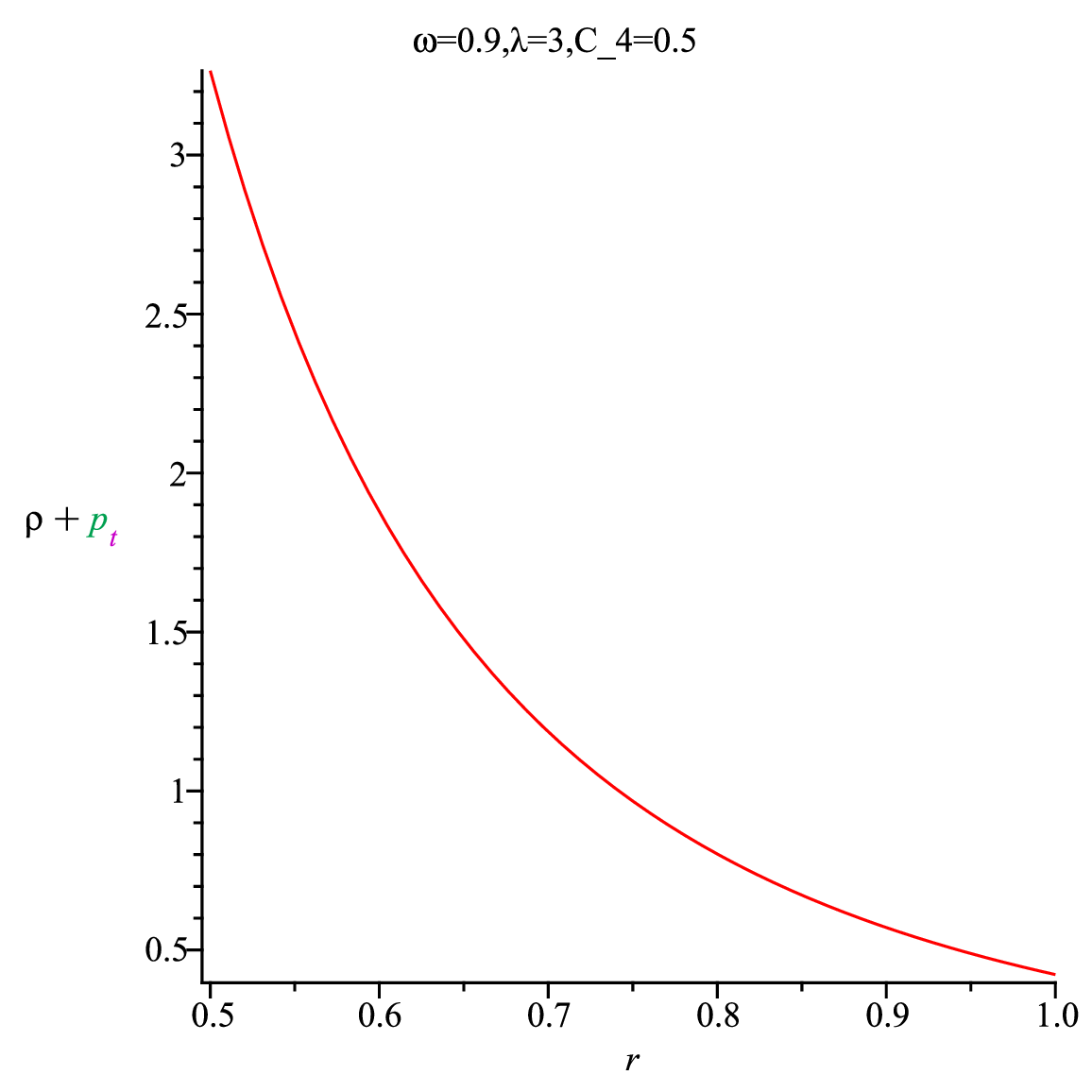}
\includegraphics[width=0.32\textwidth]{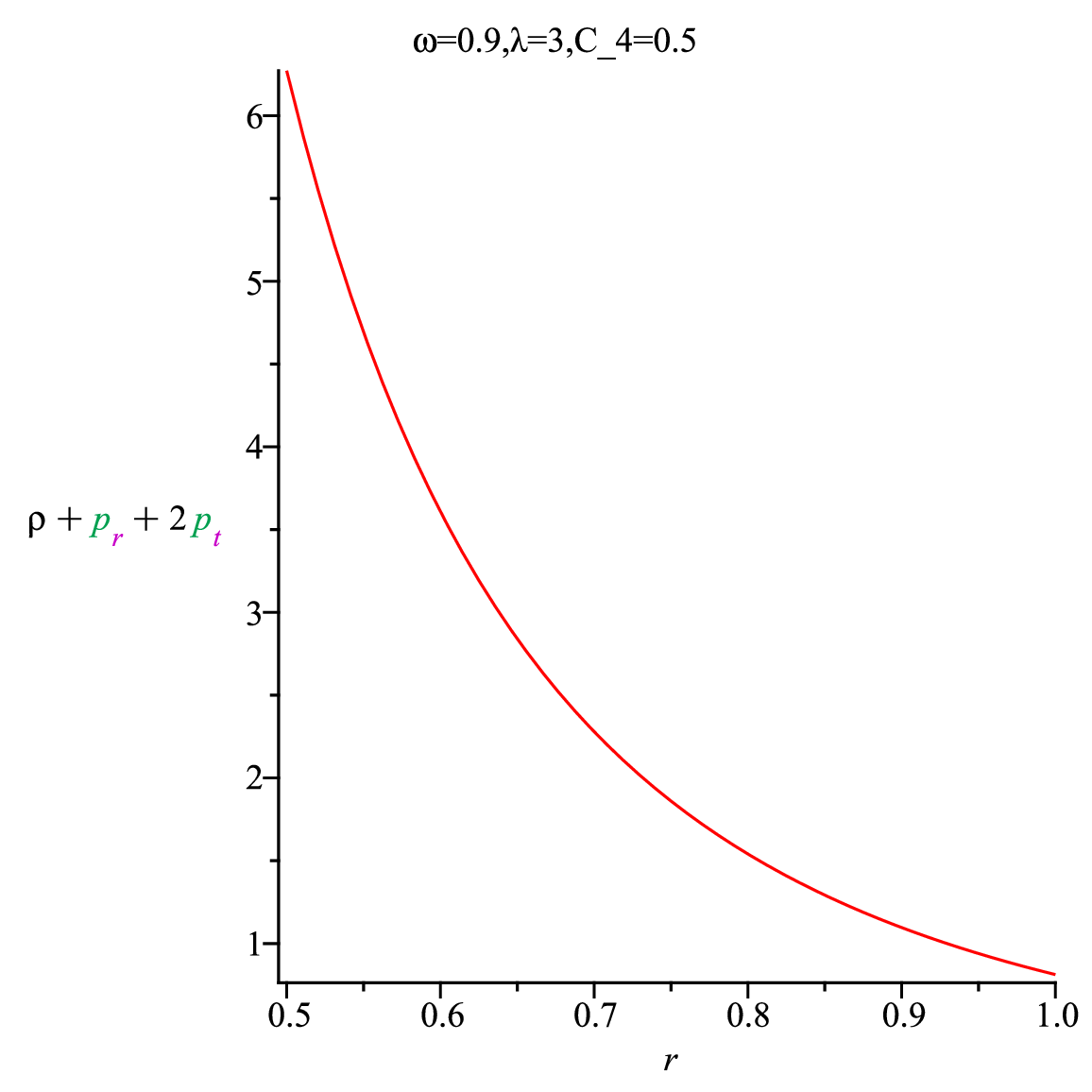}
\caption{ (  left) Variation of $\rho+p_r$     is  shown against $r$.
  ( middle)   Variation of $\rho+p_t$     is  shown against $r$.
  (  right)  Variation of $\rho+p_r+2p_t
  $     is  shown against $r$.}
\end{figure}

 \begin{figure}[thbp]
\centering
\includegraphics[width=0.4\textwidth]{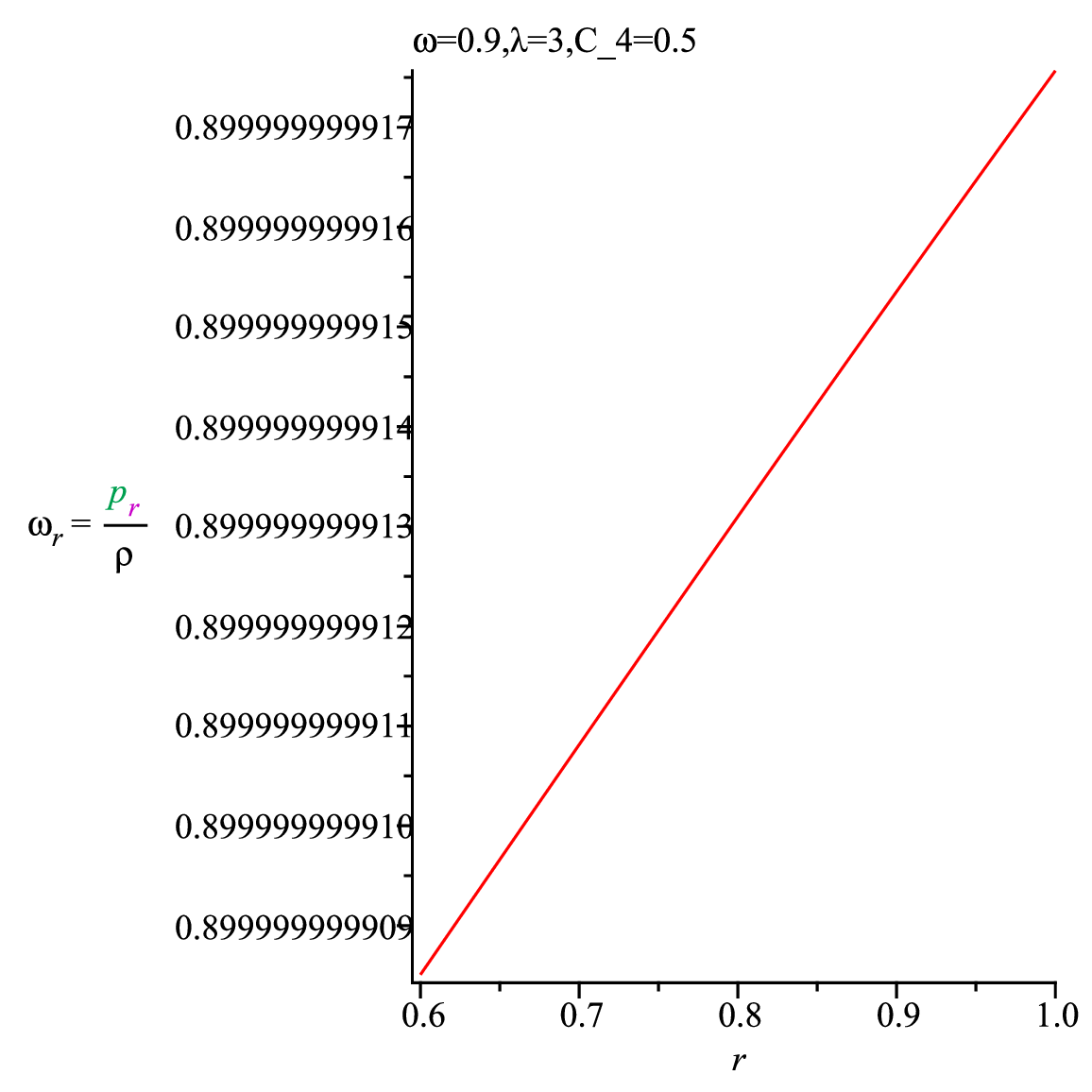}
\includegraphics[width=0.4\textwidth]{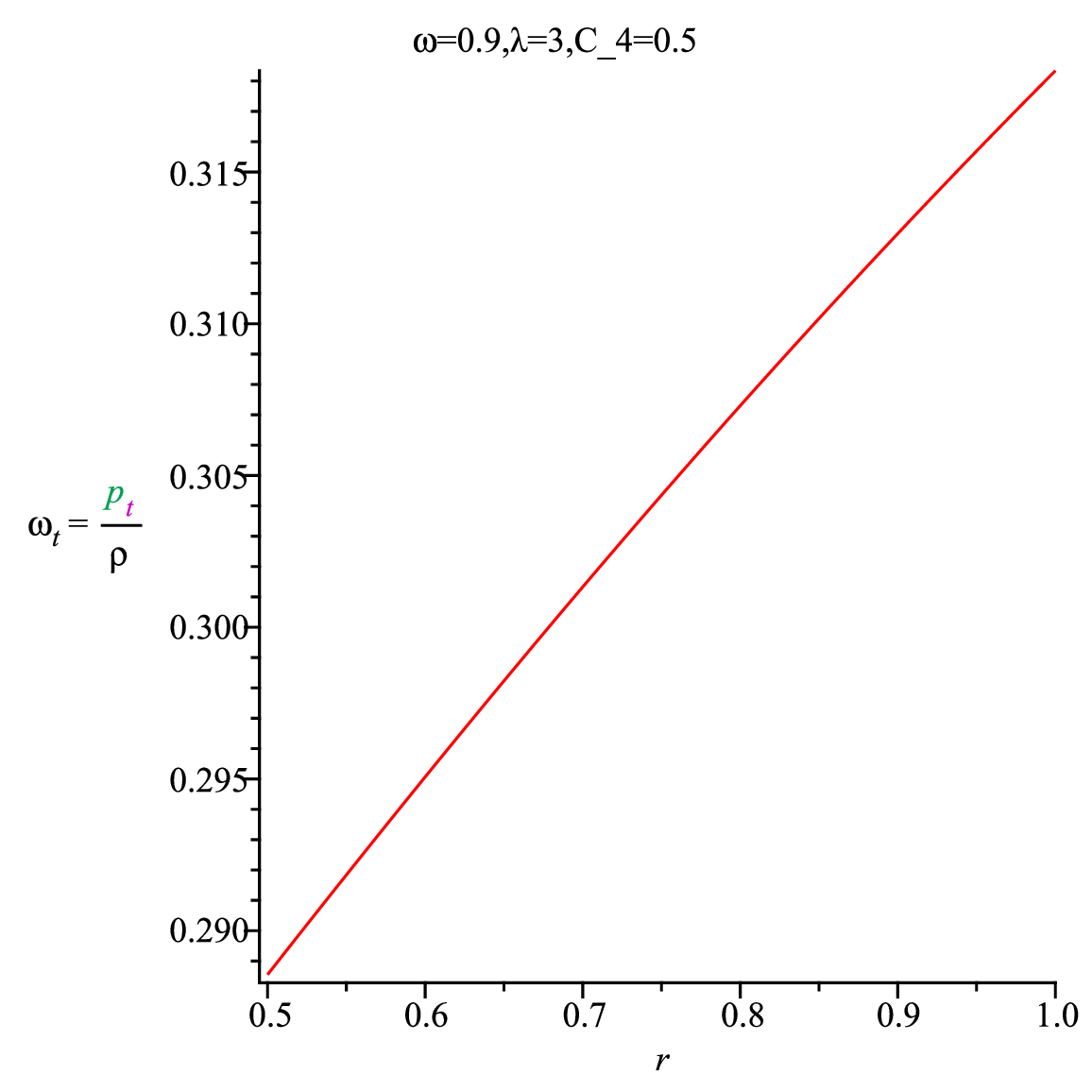}
\caption{ (  left) Variation of $\rho+p_r$     is  shown against $r$.
  ( middle)   Variation of $\rho+p_t$     is  shown against $r$.
  (  right)  Variation of $\rho+p_r+2p_t
  $     is  shown against $r$.}
\end{figure}

 \begin{figure}[thbp]
\centering
\includegraphics[width=0.32\textwidth]{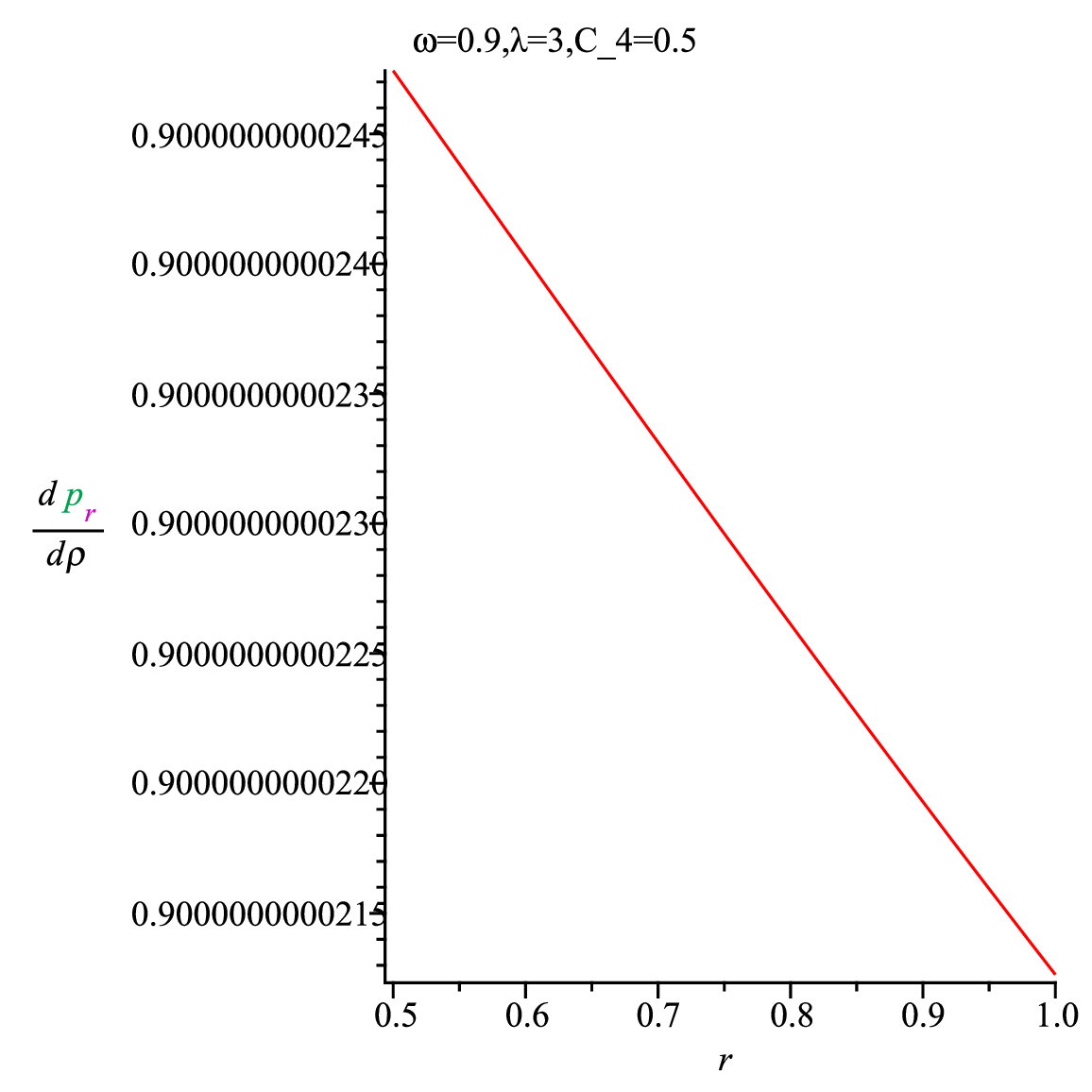}
\includegraphics[width=0.32\textwidth]{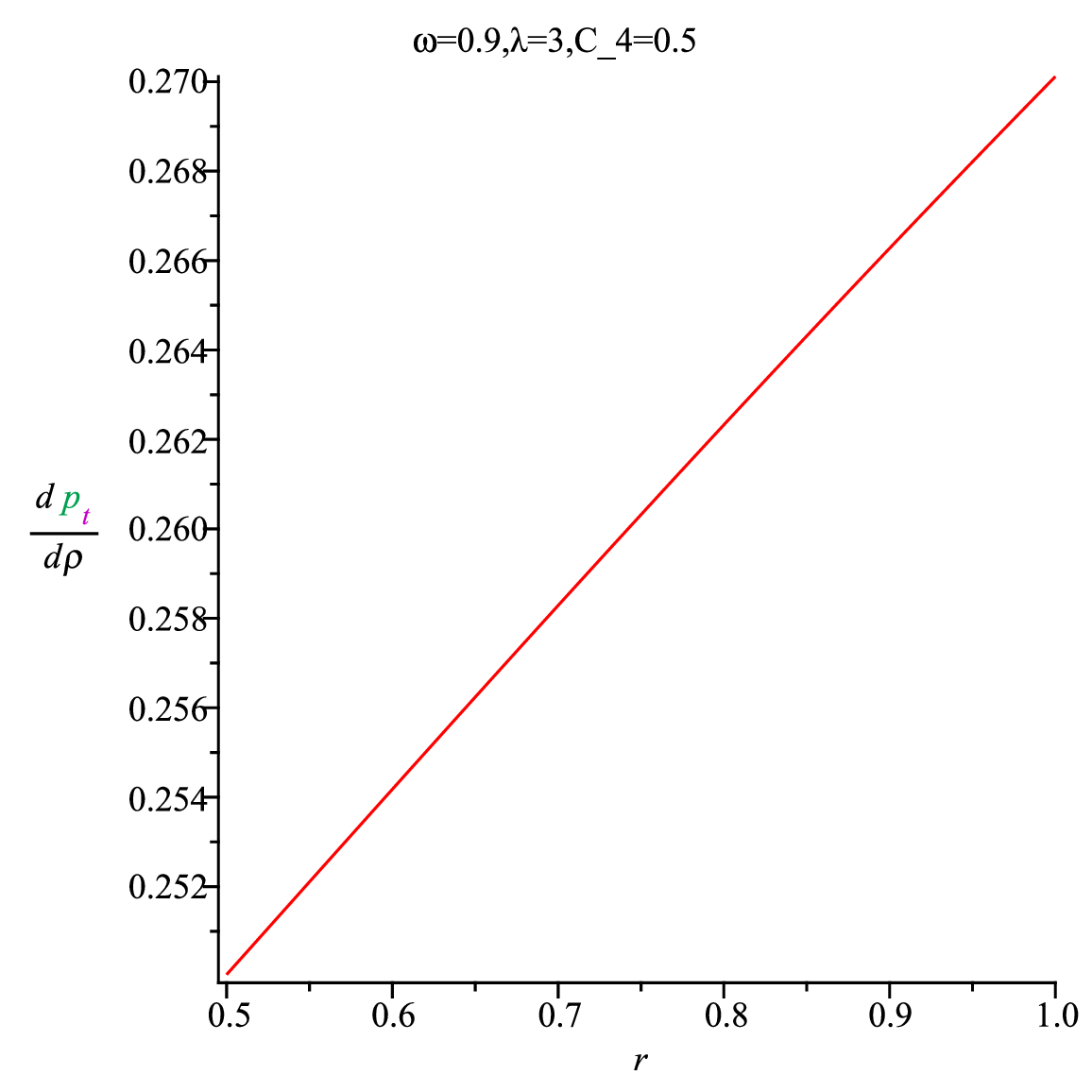}
\includegraphics[width=0.32\textwidth]{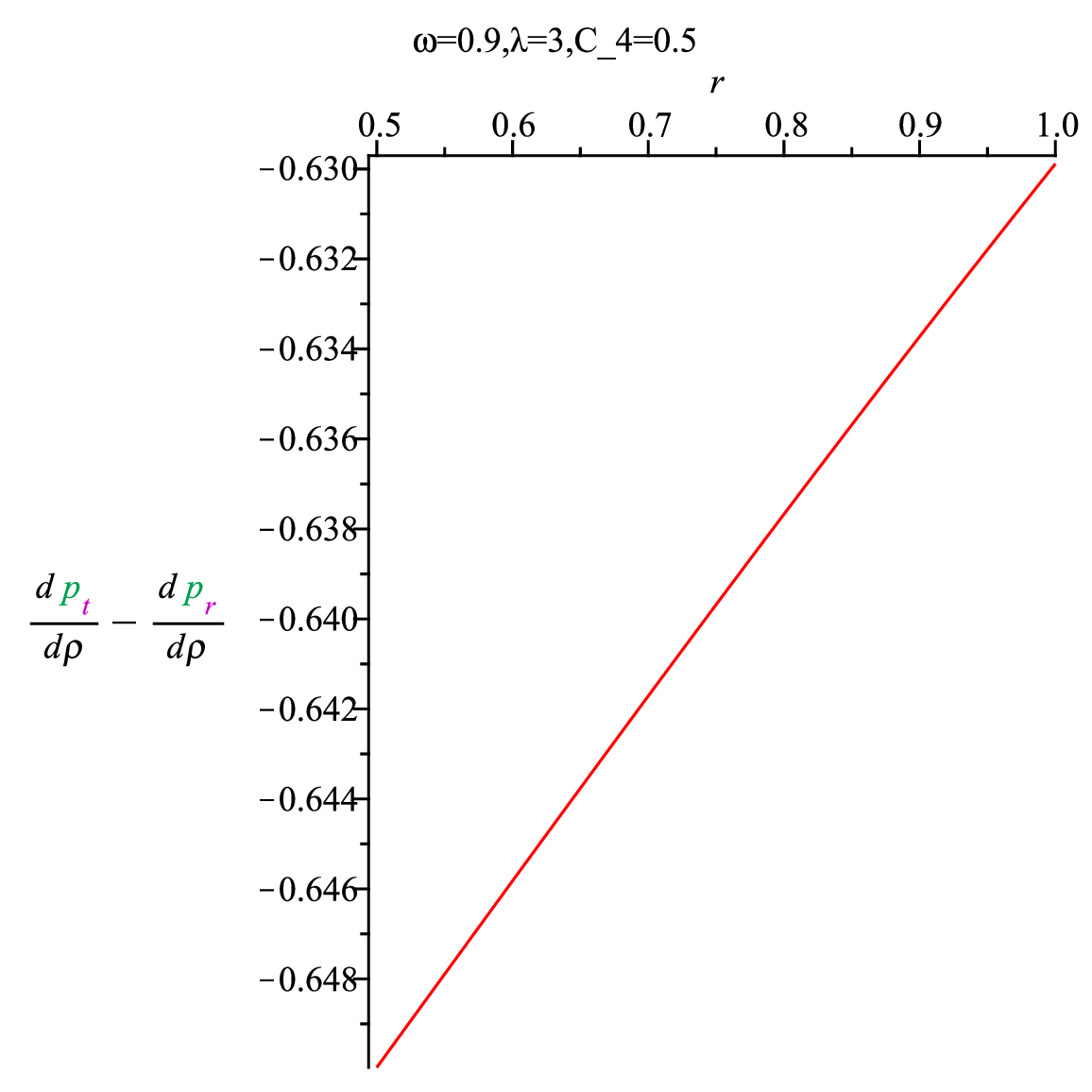}
\caption{ (  left) Variation of radial sound speed    is  shown against $r$.
  ( middle)   Variation of transverse sound speed      is  shown against $r$.
  (  right)  Variation of $v_t^2-v_r^2
  $     is  shown against $r$.}
\end{figure}

\section {Concluding Remarks }
In this paper, we have developed a compact star model in $f(R,T)$
gravity which satisfy the conformal Killing vectors equations. In
this setting, we have studied in detail the $ f(R,T)$  gravity for
the case $ f(R, T ) = R + \lambda T$ with isotropic pressure $(p_r =
p_t  = p)$  as well as anisotropic pressure $(p_r \neq p_t)$.
Further, we would like to mention that a linear equation of state
for the anisotropic case has been employed. The equation of state
parameter, integration constants and parameter of the theory
$\lambda$ have been chosen arbitrarily, so that in the present
background the value of the central energy density becomes
approximately equal to the standard value of energy density for the
compact stars. The regularity as well as energy conditions for the
both solutions have been discussed in detail.

It has been found that the energy density and pressure are positive
and finite throughout interior of the stars. The constraint on the
equation of state parameters are given by $0<{\omega}_r<1$ and
$0<{\omega}_t<1$, (as shown in figure \textbf{5}) which are in
agreement with the normal matter distribution in $f(R,T)$ gravity.
In 1992 Herrera proposed the cracking concept (also known as
overturning) which determine  the stability of anisotropic star. In
our model, we have shown from figure \textbf{(6)} that radial speed
of sound is always greater than the transverse speed of sound
everywhere inside the stars due to same sign of $v^2_r -v^2_t$ .
Therefore according to cracking concept our star model is stable in
$f(R,T)$ gravity. This work can be extended by taking more general
form of the $f(R,T)$ gravity model, to discuss some other physical
properties like anisotropic parameter $\Delta$, optimality of
density and pressure and   surface red-shift.

\subsection*{Acknowledgments}
 FR would like to thank the authorities of the Inter-University Centre
for Astronomy and Astrophysics, Pune, India for providing research facilities.  FR and IHS are also
thankful to DST, Govt. of India for providing financial support
under PURSE programme and INSPIRE Fellowship respectively.

\end{document}